\documentstyle[aps,eqsecnum,preprint,amssymb]{revtex}
\tightenlines

\begin{document}
\title{On-shell T-matrices in multiple scattering}
\author{Alexander Moroz\thanks{%
electronic-mail: moroz@amolf.nl} and Adriaan Tip\thanks{%
electronic-mail: tip@amolf.nl}}
\address{FOM-Instituut voor Atoom- en Molecuulfysica,
Kruislaan 407, SJ-1098 Amsterdam,\\ The Netherlands}
\maketitle

\begin{abstract}
The transition operator $T$ for the scattering of a particle from N
potentials $V_{j}({\bf x})$ can be expanded into a series featuring the
transition operators $t_{j}$ associated with the individual potentials. For 
$V_{j}({\bf x})$ both absolutely and square integrable in ${\bf x}$, we show,
using an analytic continuation argument, that if $T$ is on-shell, i.e. in 
$<{\bf k}|T(\sigma^{2}\pm i0)|{\bf k}^{\prime }>$, 
$|{\bf k}|=|{\bf k}^{\prime}|=\sigma$, then each $t_{j}$ is 
also on-shell.
\end{abstract}

\vspace*{1.9cm}

\begin{center}
{\bf (Phys. Lett. A 235, 195-199 (1997))}
\end{center}

\newpage

\section{Introduction}

In his booklet ``Surprises in Theoretical Physics'' Peierls \cite{Peierls}
devotes a section to off-shell effects in multiple scattering. He observes
that in the single scatterer case the on-shell T-matrix does not fix the
potential uniquely and one can try to get off-shell information by looking
at multiple scattering. Indeed, the full T-matrix can be expanded in
one-scatterer ones, where in the latter also off-shell contributions are
present. However, as observed by Baqi B\'{e}g \cite{Beg} for a two-scatterer
situation this is not the case for non-overlapping potentials. Also in
periodic systems, treated by means of the KKR-method, only the on-shell single
scatterer T-matrices occur \cite{KKR}. The above approaches make use of
partial wave expansions and this raises the question whether these can be
avoided. Below we show that this is indeed the case and that an analyticity
argument can be used instead.

In order to appreciate the situation, it is useful to consider the classical
scattering of a particle by two fixed non-overlapping finite range
potentials. Originally the particle is moving freely and after leaving the
range of the first potential it moves freely again, its energy being
unchanged, until it enters the range of the second potential.
Quantum-mechanically the situation is different. The wave function extends
throughout space at all times. Nevertheless only the on-shell single
scatterer T-operators are involved and one might think that this property is
the quantum analogue of the classical situation. However, this is not the
case. Below we show that for a large class of overlapping potentials the
on-shell T-matrix property still holds.

\section{\protect\smallskip General background}

We consider the situation of a three-dimensional Schr\"{o}dinger particle in
the field of N potentials 
\begin{equation}
H={\bf p}^{2}+\sum_{j=1}^{N}V_{j}({\bf x})=H_{0}+V.  \label{eq 2.1}
\end{equation}
A typical example is $V_{j}({\bf x})=\Phi ({\bf x-x}_{j})$, where the 
${\bf x}_{j}$'s are points in space. $N$ can be infinite as in the 
lattice case.
Then $\Phi ({\bf x})$ must have sufficient decay at infinity. The 
associated
T-matrix is 
\begin{equation}
T(z)=V+V[z-H]^{-1}V,\hspace{0.6cm}
z=E+i\varepsilon =k_{0}^{2}+i\varepsilon .
\label{eq 2.2}
\end{equation}
Expanding $T(z)$ in terms of the individual scatterer T-operators 
\begin{equation}
t_{j}(z)=V_{j}+V_{j}R_{j}(z)V_{j},\hspace{0.6cm}
H_{j}=H_{0}+V_{j},%
\hspace{0.6cm}R_{j}(z)=[z-H_{j}]^{-1},  \label{eq 2.3}
\end{equation}
we obtain the series 
\begin{equation}
T\left( z\right) =\sum_{j}t_{j}(z)+\sum_{j\neq
h}t_{j}(z)R_{0}(z)t_{h}(z)+\sum_{j\neq h\neq
k}t_{j}(z)R_{0}(z)t_{h}(z)R_{0}(z)t_{k}(z)+\ldots  \label{eq 2.4}
\end{equation}
Here the free resolvent $R_{0}(z)=[z-H_{0}]^{-1}$ contains both 
$\delta$-function and Cauchy principal value contributions with the 
result that the $t_{j}$'s need not be on-shell even if $T(z)$ itself 
is on-shell. With
on-shell we mean that $|{\bf k}_{1}|=|{\bf k}_{2}|=k_{0}$ in $<{\bf k}
_{1}|T(k_{0}^{2}+i\varepsilon )|{\bf k}_{2}>$.

Below we show that for potentials that are both {\em absolutely} and 
{\em square integrable} only the on-shell parts of the individual 
T-matrices contribute
to the on-shell part of $T(z)$. In the next section we do so for
non-overlapping potentials, whereafter we treat the general case.

\section{Non-overlapping potentials}

We start off with the objects

\begin{eqnarray}
X_{\alpha }(z) &=&
<{\bf k}_{1}|t_{j}(z)\exp [i\alpha \sqrt{H_{0}}%
]R_{0}(z)t_{h}(z)|{\bf k}_{2}>_{|{\bf k}_1|=|{\bf k}_2|=k_0}
  \nonumber \\
&=& <{\bf k}_{1}|M_{j}(z)\varphi _{j}\exp [i\alpha \sqrt{H_{0}}%
]R_{0}(z)\varphi _{h}N_{h}(z)|{\bf k}_{2}>_{|{\bf k}_1|=|{\bf k}_2|=k_0},  
\nonumber \\
Y_{\alpha }(z) &=&
<{\bf k}_{1}|\exp [-i\alpha \sqrt{H_{0}}]t_{j}(z)\exp
[i\alpha \sqrt{H_{0}}]R_{0}(z)t_{h}(z)|{\bf k}_{2}>_{|{\bf k}_1|=
|{\bf k}_2|=k_0} 
 \nonumber \\
&=&\exp [-i\alpha k_{0}]X_{\alpha }(z)  \label{eq 3.1}
\end{eqnarray}
where $j\neq h$, $z=k_{0}^{2}+i\varepsilon $, $k_{0}>0$, $\varepsilon >0$, $%
\alpha >0$ , $V_{j}({\bf x})=\varphi _{j}({\bf x})^{2}$ (so $\varphi _{j}(%
{\bf x})$ is purely imaginary for those ${\bf x\ }$ for which $V_{j}({\bf x})
$ is negative, but this does not matter in the following) and 
\begin{equation}
M_{j}(z)=\varphi _{j}\{1+\varphi _{j}R_{j}(z)\varphi_{j}\},
\hspace{0.6cm}
N_{h}(z)=\{1+\varphi _{h}R_{h}(z)\varphi _{h}\}\varphi _{h}.
\label{eq 3.2}
\end{equation}
First, consider the limit as $\varepsilon \downarrow 0$ of 
\begin{equation}
X_{0}(z)=Y_{0}(z)=<{\bf k}_{1}|t_{j}(z)R_{0}(z)t_{h}(z)|{\bf k}_{2}>.
\label{eq 3.3}
\end{equation}
We assume that each $V_{j}({\bf x})$ is both absolutely and square
integrable. Thus $<{\bf k}_{1}|\varphi _{j}$ and $\varphi _{h}|{\bf k}_{2}>$
are square integrable and moreover each $\varphi _{j}R_{j}(z)\varphi _{j}$
has a limit as $\varepsilon \downarrow 0$ \cite{Simon}\cite{RS4}. Now 
\begin{eqnarray}
X_{\alpha }(z) &=&\int d{\bf k}\exp [i\alpha \sqrt{k^{2}}][z-k^{2}]^{-1}<
{\bf k}_{1}|M_{j}(z)\varphi _{j}|{\bf k}><{\bf k}|\varphi _{h}N_{h}(z)|{\bf k
}_{2}>  \nonumber \\
&=&\int d{\bf k}\exp [i\alpha \sqrt{k^{2}}][z-k^{2}]^{-1}  \nonumber \\
&&\int d{\bf x}d{\bf y}<{\bf k}_{1}|M_{j}|{\bf x}>\varphi _{j}({\bf x})<{\bf 
x}|{\bf k}><{\bf k}|{\bf y}>\varphi _{h}({\bf y})<{\bf y}|N_{h}|{\bf k}_{2}>
\nonumber \\
&=&(2\pi )^{-3}\int d{\bf k}\exp [i\alpha \sqrt{k^{2}}][z-k^{2}]^{-1} 
\nonumber \\
&&\int d{\bf x}d{\bf y}<{\bf k}_{1}|M_{j}|{\bf x}>\varphi _{j}({\bf x})\exp
[i{\bf k\cdot (x-y)]}\varphi _{h}({\bf y})<{\bf y}|N_{h}|{\bf k}_{2}> 
\nonumber \\
&=&-\pi ^{-1}\int_{-\infty }^{+\infty }dkk\exp [i\alpha \sqrt{k^{2}}
][k^{2}-z]^{-1}<{\bf k}_{1}|M_{j}\tilde{K}(k^{2}+i0)N_{h}|{\bf k}_{2}>,
\label{eq 3.4}
\end{eqnarray}
where $\tilde{K}(z)$ is defined by the kernel 
\begin{eqnarray}
<{\bf x}|\tilde{K}(z)|{\bf y}> &=&\varphi _{j}({\bf x})\frac{\exp [i\sqrt{z}|
{\bf x-y}|}{4\pi i|{\bf x-y}|}{\bf ]}\varphi _{h}({\bf y}),  \nonumber \\
<{\bf x|}\tilde{K}(k^{2}+i0)|{\bf y}> &=&\varphi _{j}({\bf x})
\frac{\exp [ik|
{\bf x-y}|}{4\pi i|{\bf x-y}|}{\bf ]}\varphi _{h}({\bf y}).  \label{eq 3.5}
\end{eqnarray}
We now suppose that $V_{j}({\bf x})$ is non-vanishing only on the bounded
set $A_{j}$ in coordinate space and that the sets $A_{j}$ and $A_{h}$, 
$j\neq h$, do not overlap, i.e. they have a minimal distance $d>0$. 
Then the
denominator in $<{\bf x}|\tilde{K}(k^{2}+i0)|{\bf y}>$ is 
$\geqslant d^{-1}$, so $\tilde{K}(k^{2}+i0)$ and 
$<{\bf k}_{1}|M_{j}\tilde{K}(k^{2}+i0)N_{h}|{\bf k}_{2}>$ can be 
continued analytically as entire functions of 
$\varsigma $, $k\rightarrow \varsigma \in {\Bbb C}$, with exponential decay
in the upper halfplane. Defining $\sqrt{\varsigma ^{2}}$ by laying a cut
along the positive real axis, we have $\exp [i\alpha \sqrt{\varsigma ^{2}}
]=\exp [i\alpha \varsigma ]$ which remains bounded in the upper halfplane.
We can now close the contour in (\ref{eq 3.4}) and pick up the residue in 
$\sqrt{z}$ with the result 
\begin{eqnarray}
X_{\alpha }(z) &=&-i\exp [i\alpha \sqrt{z}]<{\bf k}_{1}|M_{j}\tilde{K}%
(z)N_{h}|{\bf k}_{2}>=\exp [i\alpha \sqrt{z}]X_{0}(z),  \nonumber \\
Y_{\alpha }(z) &=&\exp [i\alpha (\sqrt{z}-k_{0})]X_{0}(z).  \label{eq 3.6}
\end{eqnarray}
One has 
\begin{equation}
a^{-1}\int_{0}^{a}d\alpha\, Y_{\alpha }(k_{0}^{2}+i0)= 
X_{0}(k_{0}^{2}+i0).  \label{eq 3.7}
\end{equation}
On the other hand 
\begin{eqnarray}
\lefteqn{a^{-1}\int_{0}^{a}d\alpha\, Y_{\alpha }(k_{0}^{2}+i0)=}
\nonumber \\
&&\int
<{\bf k}_{1}|a^{-1}\int_{0}^{a}d%
\alpha \exp [-i\alpha \sqrt{H_{0}}]t_{j}(k_{0}^{2}+i0)\exp 
[i\alpha \sqrt{H_{0}}]
|{\bf k}><{\bf k}|R_{0}(z)t_{h}(z)|{\bf k}_{2}> d{\bf k}.  
\label{eq 3.8}
\end{eqnarray}
But 
\begin{equation}
<{\bf k}_{1}|a^{-1}\int_{0}^{a}d\alpha \exp [-i\alpha \sqrt{H_{0}}]
t_{j}(z)\exp [i\alpha \sqrt{H_{0}}]|{\bf k}>=
\frac{\exp [ia(k-k_{0})]-1}{ia(k-k_{0})}
<{\bf k}_{1}|t_{j}(z)|{\bf k}>,  \label{eq 3.9}
\end{equation}
which vanishes for large $a$ unless $k=k_{0}$ indicating that only the
on-shell part of $t_{j}(k_{0}^{2}+i0)$ contributes to 
$X_{0}(k_{0}^{2}+i0)=\lim_{%
\varepsilon \rightarrow 0}
<{\bf k}_{1}|t_{j}(z)R_{0}(z)t_{h}(z)|{\bf k}_{2}>$.
Since $R_{0}(z)$ is diagonal in momentum representation,
the same is also true for $t_{h}$, i.e., only the on-shell part 
of $t_{h}$ contributes to $X_{0}(k_{0}^{2}+i0)$.
Further insight is provided by considering 
two nonoverlapping spherically symmetric (muffin tin) potentials 
$V_j$ and $V_h$,
one centered at the origin and the second at the point ${\bf R}$.
Then, in terms of spherical
Bessel functions and spherical harmonics, 
one has for ${\bf x}\in A_j $ and  ${\bf y}\in A_h$
\begin{eqnarray}
-\frac{\exp [ik_{0}|{\bf x-y}|}{4\pi |{\bf x-y}|}%
&=& ik_{0} \sum_{lm}k_{0}j_{l}(k_{0}x)
h_{l}^{(+)}(k_{0}y)\, Y_{l}^{m}({\bf x})Y_{l}^{m}({\bf y})\nonumber \\
&=& 
\sum_{lm;l'm'} j_l(k_{0}x) Y_{lm}({\bf x})\,
g_{lm;l'm'}(k_{0},{\bf R}) \, j_{l'}(k_{0}|{\bf y}-{\bf R}|) 
Y_{l'm'}^* ({\bf y}-{\bf R}),
\end{eqnarray}
where $g_{lm;l'm'}(k_{0},{\bf R})$ are the so called structure constant
(see, for example, \cite{LS}). Then
\begin{eqnarray}
\lefteqn{
<{\bf k}_{1}|t_{j}(k_{0}^{2}+i0)R_{0}(k_{0}^{2}+i0)t_{h}(k_{0}^{2}+i0)|%
{\bf k}_{2}>_{|{\bf k}_1|=|{\bf k}_2|=k_0} } \hspace{1cm} \nonumber \\
&&= \sum_{lm;l'm'}
g_{lm;l'm'}(k_{0},{\bf R})\, t_{lm}(k_0) t_{l'm'}(k_0),  \label{eq 3.10}
\end{eqnarray}
where $t_{lm}(k_0)=-\sin\eta_l \exp (i\eta_l)/k_0$ and
$\eta_l=\eta_l(k_0)$ is the corresponding phase-shift. 

We can repeat the procedure for $<{\bf k}
_{1}|t_{j}(z)R_{0}(z)t_{h}(z)R_{0}(z)t_{k}(z)|{\bf k}_{2}>$, 
$j\neq h\neq k$, and higher terms in the multiple-scattering
expansion (\ref{eq 2.4})  with the result that only the on-shell 
parts of
the individual T-matrices contribute. Apart from a change in the free
Green's functions $<{\bf x}|R_{0}(z)|{\bf y}>$, the same procedure can be
followed in the one and two-dimensional cases.

\section{Overlapping potentials}

The crucial point in the derivation of the previous section is the behaviour
of the operator $\tilde{K}(z)$ defined in (\ref{eq 3.5}). The class of
potentials, called Rollnik potentials, for which $\tilde{K}(z)$ makes sense
has been profoundly studied in the literature \cite{Simon}\cite{Weinberg}.
The absolutely and square integrable potentials are members of
this class. Here we need some subtle properties of $\tilde{K}(z)$, which
were obtained by Blauw and Tip \cite{BlauwTip}, who started off from its
time-dependent counterpart $K(t)=\varphi _{j}\exp [-iH_{0}t]\varphi _{h}$.
In the present situation (for the three-dimensionsional case) it was found
that $\tilde{K}(z)$ is analytic outside the positive real axis and that $%
\tilde{K}(k_{0}^{2}\pm i\varepsilon )$, $k>0$, have the limits $\tilde{K}%
(k_{0}^{2}\pm i0)$: 
\begin{equation}
<{\bf x}|\tilde{K}(k_{0}^{2}\pm i0)|{\bf y}>=\varphi _{j}({\bf x})\frac{\exp
[\pm ik|{\bf x-y}|}{4\pi i|{\bf x-y}|}\varphi _{h}({\bf y}).  \label{eq 4.2}
\end{equation}
In addition, 
\begin{eqnarray}
||\tilde{K}(z)||_{4} &=&[{\rm tr}\{\tilde{K}(z)^{*}\tilde{K}%
(z)\}^{2}]^{1/4}\leq c<\infty ,
\hspace{0.6cm}\forall z\in {\Bbb C},  \nonumber \\
\int_{0}^{\infty }dk^{2}||\tilde{K}(k^{2}\pm i0)||_{4}^{r} &<&\infty,
\hspace{0.6cm} r>4, 
\nonumber \\
||\tilde{K}(k^{2}\pm i0)||_{4} &\rightarrow &0,
\hspace{0.6cm} (k\rightarrow \infty).
\label{eq 4.3}
\end{eqnarray}
The first of these statements implies that $\tilde{K}(z)$ is a compact
operator. Using H\"{o}lder's inequality, it follows from the second that 
$X_{\alpha }(z)$ exists and, translating the above properties to those of 
$L(k)$ defined earlier, we conclude once more that the contour can be closed.
Now the remainder of the derivation in section III can be repeated without
change. However, (\ref{eq 3.10}) looses its meaning. Also the results of 
\cite{BlauwTip}, are not available in the one and two-dimensional cases. On
the other hand one expects that the conditions on the potentials can be
relaxed.

\section{Discussion}

Using an analytic continuation argument, we have shown that for a large class
of potentials the single scatterer T-matrices in a multiple scattering
expansion can be evaluated on the energy shell. For the individual potential 
$V_{j}$ it suffices that it is both integrable and square integrable. This
allows local Coulomb singularities and a behaviour at infinity $\thicksim
x^{-\alpha }$, $\alpha >3$ (if there is an infinity of potentials
such as on a
lattice, further requirements are needed to avoid a catastrophic building up
of tail contributions). An earlier result on overlapping potentials is that
of Faulkner and Stocks \cite{Faulkner}, who considered overlapping muffin
tin potentials. Note further that the present result strongly depends on the
dimension of the system through the specific form of the free resolvent
entering into the definition of the operator $\tilde{K}(z)$.

Turning back to Peierls' discussion, we conclude that quite generally it is
impossible to abstract off-shell T-matrix information from multiple
scattering. The situation is different if off-shell matrix elements of the $%
{\it \ }{\em density}$ {\em operator} are observed. Then off-shell T-matrix
elements link the on-shell part of the density operator at the initial time
with off-shell ones at later times. Second, if we consider three identical
particles interacting through identical pair potentials, the Fadeev
equations come into play and also here off-shell elements of the individual
T-operators appear.

\acknowledgments

This work is part of the research programme of the Stichting voor
Fundamenteel Onderzoek der Materie (Foundation for Fundamental Research on
Matter) and was made possible by financial support from the Nederlandse
Organisatie voor Wetenschappelijk Onderzoek (Netherlands Organization for
Scientific Research).

\end{document}